\documentstyle[12pt]{article}
\newcommand{\be}{\begin{eqnarray}}
\newcommand{\ee}{\end{eqnarray}}
\newcommand{\ba}{\begin{array}}
\newcommand{\ea}{\end{array}}
\newcommand{\half}{{\textstyle{\frac{1}{2}}}}
\newcommand{\partialslash}{\partial\hspace{-.5em}/\hspace{.15em}}
\newcommand{\Pslash}{P\hspace{-.5em}/\hspace{.15em}}

\newcommand{\kslash}{k\hspace{-.5em}/\hspace{.15em}}
\newcommand{\bfk}{{\bf k}}
\newcommand{\kint}{\int_{\Lambda_3}\!\frac{d^4 k}{(2\pi)^4}}

\begin{document}
\begin{flushright}
JINR preprint E2-96-482, Dubna, 1996.
\end{flushright}

\begin{center}
{\large\bf The pseudoscalar and vector excited mesons \\
in the $U(3)*U(3)$ chiral model } \\[1.5cm]
{\large\bf M.K.\ Volkov}$^{\rm 1}$ \\[0.4cm]
{\em Bogoliubov Laboratory of Theoretical Physics \\
Joint Institute for Nuclear Research \\
141980, Dubna, Moscow region, Russia}
\\[0.7cm]
\end{center}
\vspace{1cm}
\begin{abstract}
\noindent
 A chiral $U(3)*U(3)$ Lagrangian containing, besides the usual meson
fields, their first radial excitations  is constructed. The
Lagrangian is derived by bosonization of the Nambu--Jona-Lasinio quark
model with separable non-local interactions, with form factors
corresponding to 3--dimensional ground and excited state wave
functions. The spontaneous breaking of chiral symmetry
is governed by the NJL gap equation. The first radial excitations
of the pions, kaons and vector
meson nonet are described with the help of five different form factors.
Each form factor contains only one arbitrary parameter. The masses of
these meson states and the weak decay constants $F_{\pi'}$, $F_K$ and
$F_{K'}$ are calculated.\\

\end{abstract}
\vspace{1cm}
%
\vfill
\rule{5cm}{.15mm}
\\
\noindent
{\footnotesize $^{\rm 1}$ E-mail: volkov@thsun1.jinr.dubna.su} \\
\newpage
\section{Introduction}
Investigation of the radial excitations of the light mesons is of great
interest in hadronic physics. So far there are the questions connected
with the experimental and theoretical descriptions of the
radial excitations of pseudoscalar mesons. For instance, the $\pi'$
meson with the mass $(1300  \pm 100 ) MeV$ is usually identified as the
first radial excitation of the pion \cite{Rev_96}. However, indications of
a light resonance in diffractive production of $3\pi$--states have
lead to speculations that the mass of the $\pi'$ may be considerably
lower, at $\sim 750 \,{\rm MeV}$ \cite{ivanshin_93}. So far there are no
experimental data concerning the excited states of the kaons \cite{Rev_96}.
In this paper we try to describe the masses of the excited pions, kaons
and vector mesons  and the weak decay constants of pseudoscalar mesons
in the framework of the $U(3)*U(3)$ chiral model.
\par
A theoretical description of radially excited pions poses some
interesting challenges. The physics of normal pions is completely
governed by the spontaneous breaking of chiral symmetry. A convenient
way to derive the properties of soft pions is by way of an effective
Lagrangian based on a non-linear realization of chiral symmetry
\cite{CCWZ69}. When attempting to introduce higher resonances to
extend the effective Lagrangian description to higher energies, one
must ensure that the introduction of new degrees of freedom does not
spoil the low--energy theorems for pions, which are universal
consequences of chiral symmetry.
\par
A useful guideline in the construction of effective meson Lagrangians
is the Nambu--Jona-Lasinio (NJL) model, which describes the
spontaneous breaking of chiral symmetry at quark level using a
four--fermion interaction \cite{volkov_83,volk_86,ebert_86}. The
bosonization of this model and the derivative expansion
of the resulting fermion
determinant reproduce the Lagrangian of the linear sigma model, which
embodies the physics of soft pions as well as higher--derivative
terms.  With appropriate couplings the model allows to derive also a
Lagrangian for vector and axial--vector mesons.
This not only gives the correct
structure of the terms of the Lagrangian as required by chiral
symmetry, but also quantitative predictions for the
coefficients, such as $F_\pi$, $F_K$, $g_\pi$, $g_\rho$, {\em etc.},
which are in
good agreement with phenomenology. One may therefore hope that a
suitable generalization of the NJL--model may provide a means for
deriving an effective Lagrangian including also the excited mesons.
\par
When extending the NJL model to describe radial excitations of
mesons, one has to introduce non-local (finite--range) four--fermion
interactions.  Many non-local generalizations of the NJL model have
been proposed, using either covariant--euclidean \cite{roberts_88} or
instantaneous (potential--type) \cite{leyaouanc_84,pervushin_90}
effective quark interactions.  These models generally require bilocal
meson fields for bosonization, which makes it difficult to perform a
consistent derivative expansion leading to an effective
Lagrangian.
A simple alternative is the use of separable
quark interactions. There are a number of advantages of working with
such a scheme. First, separable interactions can be bosonized by
introducing local meson fields, just as the usual
NJL--model. One can thus derive an effective meson
Lagrangian directly in terms of local
fields and their derivatives. Second, separable interactions allow
one to introduce a limited number of excited states and only in a
given channel. An interesting method for describing excited meson
states in this approximation was proposed in \cite{andrianov_93}.
Furthermore, the separable interaction can be defined in Minkowski
space in a 3--dimensional (yet covariant) way, with form factors
depending only on the part of the quark--antiquark relative momentum
transverse to the meson momentum
\cite{pervushin_90,kalinovsky_89,volk_96}.
This is essential for a correct description of excited states, since
it ensures the absence of spurious relative--time excitations
\cite{feynman_71}.  Finally, as we have shown \cite{volk_96},
the form factors
defining the separable interaction can be chosen so that
the gap equation of the generalized NJL--model coincides with the one
of the usual NJL--model, whose solution is a constant
(momentum--independent) dynamical quark mass. Thus, in this approach
it is possible to describe radially excited mesons above the usual
NJL vacuum. Aside from the technical simplification the latter means
that the separable generalization contains all the successful
quantitative results of the usual NJL model.
\par
In the our previous paper \cite{volk_96} the theoretical foundations
for the choice of the pion-quark form factors in a simple extension
of the NJL model to the non-local quark interaction were discussed.
It was shown that we can choose these form factors such that
the gap equation conserves the usual form and gives the solution
with a constant constituent quark mass.
The quark condensate also does not change after including
the excited states in the model, because the tadpole connected
with the excited scalar field is equal to zero (the quark loop
with the one excited scalar vertex - vertex with form factor).
\par
Now we shall use these form factors for describing the first
excited states of the pseudoscalar and vector meson nonets
in the framework of the more realistic $U(3)*U(3)$ chiral
model \cite{volkov_83,volk_86,ebert_86}. We shall take into
account the connections of the scalar and vector coupling
constants which have appeared in this model and the additional
renormalization of the pseudoscalar fields connected with
the pseudoscalar -- axial--vector transitions.
For simplicity, we shall suppose that the masses of the $up$
and $down$ quarks are equal to each other and shall take into
account only the mass difference between ($up$, $down$)
and $strange$ quarks ($m_u$ and $m_s$). Then we have
in this model the five basic parameters: $m_u$, $m_s$,
$\Lambda_3$ (3-dimensional cut-off parameter), $G_1$
and $G_2$ (the four--quark coupling constants for
the scalar--pseudoscalar coupling ($G_1$) and for the
vector -- axial--vector coupling ($G_2$)). For the definition
of these parameters we shall use the experimental values: the
pion decay constant $F_\pi = 93 MeV$, the $\rho$--meson decay
constant $g_\rho \approx 6.14$ ($\frac{g_\rho^2}{4\pi}
\approx 3$), the pion mass $M_\pi \approx 140 MeV$,
$\rho$--meson mass $M_\rho = 770 MeV$ and the kaon mass
$M_K \approx 495 MeV$. Using these five parameters we can
describe the masses of the four meson nonets (pseudoscalar,
vector, scalar and axial--vector)
\footnote{The mass formulae for the axial--vector mesons and ,
especially, for the scalar mesons give only qualitative
results ($20-30{\%}$ accuracy).} and all the meson coupling
constants describing the strong interactions of
the meson with each other and with the quarks.
\par
For the investigation of the excited states of the mesons it
is necessary to consider the non-local four--quark interactions.
We have shown that for description of the excited states of pions,
kaons and vector meson nonet we have to use
five different form factors in the effective
four--quark interactions. Each form factor contains only one
arbitrary parameter.
We have calculated also
the weak decay constants $F_{\pi'}$, $F_{K}$
and $F_{K'}$. In our next work we are going to calculate
the decay widths of the excited mesons and to describe
the excited states of the $\eta$ and $\eta'$ mesons.
\footnote{ Remind, one more
additional parameter , connected with the gluon anomaly,
was used in the usual NJL model, when we described the ground
states of the $\eta$ and $\eta'$ mesons ($U(1)$ problem)
\cite{volk_86}. \\
The problem of the radial excitations of the light mesons,
including the $\eta$ and $\eta'$, in the framework of
the potential model was discussed in works \cite{ger_85}.}
\par
In section 2, we introduce the effective quark interaction
in the separable approximation and describe its bosonization.
We discuss the choice of the form factors
necessary to describe the excited states of the
pseudoscalar and vector meson nonets. In section 3, we derive
the effective Lagrangian for the pseudoscalar mesons, and
perform the diagonalization leading to the physical meson
ground and excited states. In section 4,
we perform it for the vector mesons. In section 5, we fix the
parameters of the model and evaluate the masses of the ground and
excited meson states and weak decay constants $F_{\pi}$, $F_{\pi'}$,
$F_K$ and $F_{K'}$. In section 6, we discuss the obtained results.
\section{$U(3)*U(3)$ chiral Lagrangian with the excited meson
states }
In the usual $U(3)*U(3)$ NJL model a local (current--current)
effective quark interaction is used
\be
L [\bar q, q] =
\int d^4 x \, \bar q (x) \left( i \partialslash - m^0 \right)
q (x) \; + \; L_{\rm int} ,
\label{L_NJL}
\ee
\be
L_{\rm int} &=& \int d^4 x [ \frac{G_1}{2}
( j_S^a (x) j_S^a (x) +
j_P^a (x) j_P^a (x) )   \nonumber \\
&-& \frac{G_2}{2}
( j_V^a (x) j_V^a (x) +
j_A^a (x) j_A^a (x) )  ] ,
\label{L_int}
\ee
where $m^0$ is the current quark mass matrix. We suppose that
$m_u^0 \approx m_d^0$. \\
$j^a_{S,P,V,A} (x)$ denote, respectively, the scalar,
pseudoscalar, vector and axial--vector currents of the
quark field ($U(3)$--flavor),
\be
j^a_S (x) &=& \bar q (x) \lambda^a q (x), \hspace{2cm}
j^a_P (x) = \bar q (x) i\gamma_5 \lambda^a q (x) , \nonumber \\
j^{a,\mu}_V (x) &=& \bar q (x) \gamma^{\mu} \lambda^a q (x),
\hspace{1.5cm}
j^{a,\mu}_A (x) = \bar q (x) \gamma_5 \gamma^{\mu} \lambda^a q (x).
\label{j_def}
\ee
Here $\lambda^a$ are the Gell-Mann matrices, $0 \leq a \leq 8$.
The model can be bosonized in the standard way by representing the
4--fermion interaction as a Gaussian functional integral over
scalar, pseudoscalar,vector and axial--vector meson fields
\cite{volkov_83,volk_86,ebert_86}.
The effective meson Lagrangian, which is obtained by integration
over the quark fields, is expressed in terms of local meson fields.
By expanding the quark determinant in derivatives of the
local meson fields one then derives the chiral meson Lagrangian.
\par
The Lagrangian (\ref{L_int}) describes only ground--state
mesons. To include excited states, one has to introduce effective
quark interactions with a finite range.  In general, such
interactions require bilocal meson fields for bosonization
\cite{roberts_88,pervushin_90}. A possibility to avoid this
complication is the use of a separable interaction, which is still
of current--current form, eq.(\ref{L_int}), but allows for non-local
vertices (form factors) in the definition of the quark currents,
eqs.(\ref{j_def}),
\be
\tilde{L}_{\rm int} =
\int d^4 x \sum_{i = 1}^N \left[ \frac{G_1}{2}
\left[ j_{S,i}^a (x) j_{S,i}^a (x) +
j_{P,i}^a (x) j_{P,i}^a (x) \right]  \right. \nonumber \\
\left. - \frac{G_2}{2}  \left[ j_{V,i}^a (x) j_{V,i}^a (x) +
j_{A,i}^a (x) j_{A,i}^a (x) \right]  \right] ,
\label{int_sep}
\ee
\be
j^a_{S, i} (x) &=& \int d^4 x_1 \int d^4 x_2 \;
\bar q (x_1 ) F^a_{S, i} (x; x_1, x_2 ) q (x_2 ),
\label{j_S} \\
j^a_{P, i} (x) &=& \int d^4 x_1 \int d^4 x_2 \;
\bar q (x_1 ) F^a_{P, i} (x; x_1, x_2 ) q (x_2 ),
\label{j_P} \\
j^{a,\mu}_{V, i} (x) &=& \int d^4 x_1 \int d^4 x_2 \;
\bar q (x_1 ) F^{a,\mu}_{V, i} (x; x_1, x_2 ) q (x_2 ),
\label{j_V} \\
j^{a,\mu}_{A, i} (x) &=& \int d^4 x_1 \int d^4 x_2 \;
\bar q (x_1 ) F^{a,\mu}_{A, i} (x; x_1, x_2 ) q (x_2 ),
\label{j_A}
\ee
Here, $F^{a,\mu}_{U, i}(x; x_1, x_2 )$, \,
$i = 1, \ldots N$, denote a set of non-local scalar,
pseudoscalar, vec\-tor and axial--vector quark ver\-tices
(in general momentum-- and spin--dependent),
which will be specified below. Upon bosonization
we obtain
\be
L_{\rm bos}(\bar q, q; \sigma, \phi, P, A) = \int d^4 x_1
\int d^4 x_2~ \bar q (x_1 ) [ \left( i \partialslash_{x_2}
- m^0 \right) \delta (x_1 - x_2 )      \nonumber \\
+ \int d^4 x  \sum_{i = 1}^N
\left( \sigma^a_i (x) F^a_{\sigma , i} (x; x_1, x_2 ) +
\phi_i^a (x) F_{\phi , i}^a (x; x_1, x_2)  \right. \nonumber \\
\left.
+ V_i^{a,\mu} (x) F_{V , i}^{a,\mu} (x; x_1, x_2) +
A_i^{a,\mu} (x) F_{A , i}^{a,\mu} (x; x_1, x_2) \right) ] q (x_2 )
\nonumber \\
- \int d^4 x \sum_{i = 1}^N
\left[ \frac{1}{2G_1} \left( \sigma_i^{a\, 2} (x) +
\phi_i^{a\, 2} (x) \right)
- \frac{1}{2G_2} \left( V_i^{a,\mu\, 2} (x) + A_i^{a,\mu\, 2} (x)
\right) \right].
\label{L_sep}
\ee
This Lagrangian describes a system of local meson fields,
$\sigma_i^a (x)$, $\phi_i^a (x)$, $V^{a,\mu}_i (x)$, $A^{a,\mu}_i (x)$,
$i = 1, \ldots N$, which interact with
the quarks through non-local vertices. These fields
are not yet to be associated with physical particles ,
which will be obtained after determining the vacuum and
diagonalizing the effective meson Lagrangian.
\par
In order to describe the first radial excitations of mesons (N = 2),
we take the form factors in the form (see \cite{volk_96} )
\be
F^a_{\sigma , 2} ({\bf k}) &=& \lambda^a f_a^P ({\bf k}),
;\;\;\;\;
F^a_{\phi , 2} ({\bf k}) = i \gamma_5 \lambda^a f_a^P ({\bf k}),
\nonumber \\
F^{a,\mu}_{V , 2} ({\bf k}) &=& \gamma^\mu \lambda^a f_a^V ({\bf k}),
;\;\;\;\;
F^{a,\mu}_{A , 2} ({\bf k}) = \gamma_5 \gamma^\mu \lambda^a
f_a^V ({\bf k}),
\label{ffs}
\ee
\be
f_a^U ({\bf k}) = c_a^U ( 1 + d_a {\bf k}^2 ).
\label{ff}
\ee
We consider here the form factors in the momentum space and in
the rest frame of the mesons (${\bf P}_{meson}$ = 0. $k$ and
$P$ are the relative and total momentum of the quark-antiquark
pair.). For the ground states of the mesons the functions
$f_a^{U,0} ({\bf k})$ = 1.
\par
After integrating over the quark fields in
eq.(\ref{L_sep}), one obtains the effective Lagrangian of the
$\sigma_1^a , \sigma_2^a , \phi_1^a,  \phi_2^a, V_1^{a,\mu},
V_2^{a,\mu}, A_1^{a,\mu}$ and $ A_2^{a,\mu}$ fields.
($u_1 = u, u_2 = \bar u$)
\be
L(\sigma', \phi, V, A, \bar\sigma, \bar\phi, \bar V, \bar A) =\;\;\;\;\;\;\;~~~~~~~~~~
\nonumber \\
- \frac{1}{2 G_1} (\sigma_a^{'2} + \phi_a^2 + \bar\sigma_a^2 +
\bar\phi_a^2 )
+ \frac{1}{2 G_2} (V_a^2 + A_a^2 + \bar V_a^2 + \bar A_a^2 )
\nonumber \\
- i N_c \; {\rm Tr}\, \log [ i \partialslash - m^0 + (\sigma'_a
+ i \gamma_5  \phi_a +\gamma_\mu V^\mu_a +
\gamma_5 \gamma_\mu A^\mu_a   \nonumber \\
+ (\bar\sigma_a + i \gamma_5 \bar\phi_a) f_a^P + (\gamma_\mu \bar V^\mu_a +
\gamma_5 \gamma_\mu \bar A^\mu_a ) f_a^V ) \lambda^a ]
\label{12}
\ee
\par
Now let us remind how we fix the basic parameters in the usual
NJL model without the excited state of mesons \cite{volk_86}.
\par
Firstly, define the vacuum expectation of the $\sigma'_a$ fields
\be
<\frac{\delta L}{\delta\sigma'_a}>_0 &=& - i N_c \; {\rm tr} \kint
\frac{1}{( \rlap/k - m^0 + <\sigma'_a>_0 )}
- \frac{<\sigma'_a>_0}{G_1} \; = \; 0 .
\label{gap_1}
\ee
Introduce the new sigma fields whose vacuum expectations are
equal to zero
\be
\sigma_a = \sigma'_a - <\sigma'_a>_0
\label{sigma}
\ee
and redefine the quark masses
\be
m_a = m_a^0 - <\sigma'_a>.
\label{m^0}
\ee
Then eq. (\ref{gap_1}) can be rewritten in the form of the usual
gap equation
\be
m_i = m_i^0 + 8 G_1 m_i I_1 (m_i),  \;\;\;\;\;\; (i = u, d, s)
\label{gap}
\ee
where
\be
I_n (m_i) = -i N_c \; \kint \frac{1}{(m_i^2 - k^2)^n}
\label{I_n}
\ee
and $m_i$ are the constituent quark masses.
\par
In order to obtain the correct coefficients of kinetic terms
of the mesons in the one-quark-loop approximation, we have to
make the renormalization of
the meson fields in eq. (12)
\be
\sigma_a = g_{\sigma}^a \sigma_a^r, \;\;\;
\phi_a = g_{\sigma}^a \phi_a^r, \;\;\;
V^{\mu}_a = g_V^a V^{\mu,r}_a, \;\;\; A^{\mu}_a = g_V^a
A^{\mu,r}_a,
\label{ren}
\ee
where
\be
g_{\sigma}^{a_{i,j}} = [4 I_2 (m_i, m_j)]^{-\frac{1}{2}}, \;\;\;
I_2 (m_i, m_j) = -i N_c \; \kint \frac{1}{(m_i^2 - k^2)(m_j^2 -
k^2)} ,
\label{g_sigma}
\ee
\be
g_V^a = \sqrt{6} g_{\sigma}^a .
\label{g_V}
\ee
After taking into account the pseudoscalar -- axial--vector
transitions ($\phi_a \rightarrow A_a$), the additional
renormalization of the pseudoscalar fields appears
\be
g^a_{\phi} = Z_a^{-\frac{1}{2}} g_{\sigma}^a,
\label{g_phi}
\ee
where $Z_{\pi} = 1 - \frac{6 m^2_u}{M^2_{a_1}} \approx 0.7$ for
pions. ($M_{a_1} = 1.23 GeV$ is the mass of the axial-vector
$a_1$ meson, \cite{Rev_96}, $m_u = 280 MeV$ \cite{volk_86} ).
We shall assume that all $Z_a \approx Z_\pi \approx 0.7$.
\par
After these renormalizations the part of the Lagrangian (12)
describing the ground states of mesons takes the form
\be
L(\sigma, \phi, V, A)
= - \frac{1}{2 G_1} (g^{a 2}_\sigma \sigma_a^2 + g^{a 2}_{\phi}
\phi_a^2 ) +
\frac{g_V^{a 2}}{2 G_2} ( V_a^2 + A_a^2 )   \nonumber \\
- i N_c~{\rm Tr}~\log \left[ i \partialslash - m^0 +
\left( g^a_\sigma \sigma_a + i \gamma_5 g^a_{\phi} \phi_a
+\frac{g_V^a}{2} (\gamma_\mu V^\mu_a +
\gamma_5 \gamma_\mu A^\mu_a ) \right) \lambda^a \right].
\label{Lg}
\ee
For simplicity we omitted here the index $r$ of the meson fields.
\par
>From the Lagrangian (\ref{Lg}) in the one-loop approximation the
following expressions for the meson masses are obtained
\cite{volk_86}
\be
M_\pi^2 &=& g_\pi^2 \left[ \frac{1}{G_1} - 8 I_1 (m_u) \right] =
\frac{g_\pi^2}{G_1} \frac{m_u^0}{m_u}, \;\;\;\;
g_\pi^2 = \frac{1}{4 Z I_2 (m_u, m_u)},
\label{M_pi}
\ee
\be
M_K^2 = g_K^2 \left[ \frac{1}{G_1} - 4 ( I_1 (m_u) + I_1 (m_s) )
\right] + Z^{-1} (m_s - m_u)^2, \nonumber \\
g_K^2 = \frac{1}{4 Z I_2 (m_u, m_s)},\hspace{4cm}
\label{M_K}
\ee
\be
M_\rho^2 = \frac{g_\rho^2}{4 G_2} = \frac{3}{8 g_2
I_2 (m_u, m_u)},
\label{M_rho}
\ee
\be
M_\phi^2 = M_\rho^2 \frac{I_2 (m_u, m_u)}{I_2 (m_s, m_s)},
\label{M_phi}
\ee
\be
M_{K^*}^2 = M_\rho^2 \frac{I_2 (m_u, m_u)}{I_2 (m_u, m_s)} +
\frac{3}{2} (m_s - m_u)^2.
\label{M_K*}
\ee
Now let us fix our basic parameters. For that we shall use the
five experimental values \cite{volkov_83,volk_86}: \\
1) The pion decay constant $F_\pi = 93 MeV$ .\\
2) The $\rho$-meson decay constant $g_\rho \approx 6.14$. \\
Then from the Goldberger-Treimann identity we obtain
\be
m_u = F_\pi g_\pi
\label{GT}
\ee
and from eqs. (\ref{g_V}) and (\ref{g_phi}) we get
\be
g_\pi = \frac{g_\rho}{\sqrt{6 Z}}, \;\;\;\;
m_u = \frac{F_\pi g_\rho}{\sqrt{6 Z}} = 280 MeV.
\label{m_u}
\ee
>From eqs. (\ref{g_sigma}) and (\ref{g_V}) we can obtain
(see \cite{ebert_93})
\be
I_2 (m_u, m_u) = \frac{3}{2 g_\rho^2},\hspace{2cm}
\Lambda_3 = 1.03 GeV.
\label{Lambda_3}
\ee
3) $M_\pi \approx 140 MeV$. The eq. (\ref{M_pi}) gives
$G_1 = 3.48 GeV^{-2}$
(see \cite{ebert_93}). \\
4) $M_\rho = 770 MeV$. The eq. (\ref{M_rho}) gives
$G_2 = 16 GeV^{-2}$. \\
5) $M_K \approx 495 MeV$. The eq. (\ref{M_K}) gives
$m_s = 455 MeV$.  \\
After that the masses of $\eta, \eta'$ and $K^*, \phi$ mesons
can be calculated with a satisfactory accuracy.
\footnote{To calculate the masses of
the $\eta$ and $\eta'$ mesons, it is necessary to take into account
the gluon anomaly \cite{volk_86}.} It is possible also to give the
qualitative estimations for the masses of the scalar and axial--vector
mesons, using the formulae
\be
M_{A_{i,j}}^2 = M_{V_{i,j}}^2 + 6 m_i m_j, \\
M_{\sigma_{i,j}}^2 = M^2_{\phi_{i,j}} + 4 m_i m_j .
\label{M_{A,sigma}}
\ee
We can calculate the values of all the coupling constants, describing
the strong interactions of the scalar, pseudoscalar, vector and
axial--vector mesons with each other and with the quarks, and
describe all the main decays of these mesons (see \cite{volk_86}).

\section{The effective Lagrangian for the ground and excited states
of the pions and kaons}
To describe the first excited states of all the meson nonets, it
is enough to use only three different inner parameters $d_a$ in
form factors $f_a^U ({\bf k})$ (see eq. (\ref{ff}))
\be
f_{uu}^{P,V} ({\bf k}) = c_{uu}^{P,V} ( 1 + d_{uu} {\bf k}^2 ), \nonumber\\
f_{us}^{P,V} ({\bf k}) = c_{us}^{P,V} ( 1 + d_{us} {\bf k}^2 ), \nonumber\\
f_{ss}^{P,V} ({\bf k}) = c_{ss}^{P,V} ( 1 + d_{ss} {\bf k}^2 ).
\label{ffq}
\ee
Following our work \cite{volk_96} we can fix the parameters $d_{uu},
d_{us} $ and $d_{ss}$ by using the conditions
\be
I_1^{f_{uu}^P} (m_u) = 0,\;\;~~~
I_1^{f_{us}^P} (m_u) + I_1^{f_{us}^P} (m_s) = 0, \;\;~~~
I_1^{f_{ss}^P} (m_s) = 0,
\label{I_1^f}
\ee
where
\be
I_1^{f_a..f_a} (m_i) = -i N_c \;
\kint \frac{f_a..f_a}{(m_i^2 - k^2)}.
\label{I_1^ff}
\ee
The eqs. (\ref{I_1^f}) allows us to conserve the gap
equations in the form usual for the NJL model (see eqs.
(\ref{gap})), because the tadpoles
with the excited scalar external fields do not contribute to
the quark condensates and to the constituent quark masses.
\par
Using eqs. (\ref{I_1^f}) we obtain for all $d_a$ close values
\be
d_{uu} = - 1.784~ GeV^{-2},~~ d_{us} = - 1.7565~ GeV^{-2},~~
d_{ss} = - 1.727~ GeV^{-2}.
\label{d_a}
\ee
\par
Now let us consider the free part of the Lagrangian (12).
For the pseudoscalar meson we obtain
\be
L^{(2)} (\phi) &=&
\half \sum_{i, j = 1}^{2} \sum_{a = 0}^{8}
\phi_i^a (P) K_{ij}^{ab} (P) \phi_j^b (P) .
\label{L_2}
\ee
Here
\be
\sum_{a = 1}^{3} (\phi_i^a)^2 = (\pi_i^0)^2 + 2 \pi^+_i \pi^-_i,~~~
(\phi_i^4)^2 + (\phi_i^5)^2 = 2 K_i^+ K_i^-, \nonumber \\
(\phi_i^6)^2 + (\phi_i^7)^2 = 2 K_i^0 \bar K_i^0, ~~~
(\phi_i^0)^2 = (\phi_i^u)^2,~~~(\phi_i^8)^2 = (\phi_i^s)^2.
\label{phi^a}
\ee
$\phi_i^u$ and $\phi_i^s$ are the components of the $\eta$ and
$\eta'$ mesons.
The quadratic form $K_{ij}^{ab} (P)$, eq.(\ref{L_2}), is
obtained as
\be
K_{ij}^{ab} (P) \equiv \delta^{ab} K_{ij}^a (P) ,~~~~~~~~~~~~~~~~~~~~~~~~
\nonumber \\
K_{ij}^a (P) = -~\delta_{ij} \frac{1}{G_1}~~~~~~~~~~~~~~~~~~~~~~~~
\nonumber \\
-~i~ N_{\rm c} \; {\rm tr}\, \kint \left[
\frac{1}{\kslash + \half\Pslash - m_q^a}
i\gamma_5 f_i^a
\frac{1}{\kslash - \half\Pslash - m_{q'}^a} i \gamma_5 f_j^a
\right]  , \nonumber \\
f_1^a \equiv 1, \hspace{2em} f_2^a \;\; \equiv \;\; f_a^P ({\bf k}).~~~~~~~~~~
\label{K_full}
\ee
\be
m_q^a &=& m_u~~(a = 0,...,7);~~m_q^8 = m_s;  \nonumber \\
m_{q'}^a &=& m_u~~(a = 0,...,3);~~ m_{q'}^a = m_s~~ (a = 4,...,8).
\label{m_q^a}
\ee
$m_u$ and $m_s$ are the constituent quark masses ($m_u \approx m_d$).
The integral (\ref{K_full}) is evaluated by expanding in the
meson field momentum, $P$. To order $P^2$, one obtains
\be
K_{11}^a(P) &=& Z_1^a (P^2 - M_1^{a 2} ),
\hspace{2em} K_{22}^a(P) \;\; = \;\; Z_2^a (P^2 - M_2^{a 2} )
\nonumber \\
K_{12}^a(P) &=& K_{21}^a(P) \;\; = \;\;
\gamma^a (P^2 - \Delta^2 \delta_{ab}|_{b = 4,...,7}),~~
(\Delta = m_s - m_u)
\label{K_matrix}
\ee
where
\be
Z_1^a &=& 4 I_2^a Z , \hspace{2em} Z_2^a \; = \; 4 I_2^{ff a} \bar{Z},
\hspace{2em} \gamma^a \; = \; 4 I_2^{f a} Z,
\label{I_12} \\
M_1^{a 2} &=& (Z_1^a)^{-1}[\frac{1}{G_1}-4(I_1^a(m_q^a) +
I_1^a(m_{q'}^a)] +
Z^{-1} \Delta^2 \delta_{ab}|_{b = 4,...,7}~ ,
\label{M_1} \\
M_2^{a 2} &=& (Z_2^a)^{-1}[\frac{1}{G_1}-4(I_1^{ff a}(m_q^a) +
I_1^{ff a}(m_{q'}^a)] + \bar{Z}^{-1} \Delta^2 \delta_{ab}|_{b = 4,...,7}~ .
\label{M_2}
\ee
Here, $\bar{Z} = 1 - \Gamma^2~\frac{6 m^2_u}{M^2_{a_1}} \approx 1$ (see eq.
(49)),
$I_n^a, I_n^{f a}$ and $I_n^{ff a}$ denote the usual loop
integrals arising in the momentum expansion of the NJL quark
determinant, but now with zero, one or two factors
$f_a ({\bf k})$, eqs.(\ref{ffq}), in the
numerator (see (\ref{I_1^ff}) and below )
\be
I_2^{f..f a} (m_q, m_{q'}) &=& -i N_{\rm c}
\kint \frac{f_a(\bfk )..f_a(\bfk )}{(m_q^{a 2} - k^2)
(m_{q'}^{a 2} - k^2)}.
\label{I_2^ff}
\ee
The evaluation of these integrals with a 3--momentum cutoff is
described {\em e.g.}\ in ref.\cite{ebert_93}. The integral over $k_0$
is taken by contour integration, and the remaining 3--dimensional
integral is regularized by the cutoff. Only the divergent parts are
kept; all finite parts are dropped. We point out that the momentum
expansion of the quark loop integrals, eq.(\ref{K_full}), is an
essential part of this approach.  The NJL--model is understood here
as a model only for the lowest coefficients of the momentum expansion
of the quark loop, not its full momentum dependence (singularities
{\em etc.}).
$Z$ is the additional renormalization of the ground pseudoscalar meson
states taking into account the $\phi^a \rightarrow A^a$ transitions
(see eq.(\ref{g_phi})).
\par
After the renormalization of the meson fields
\be
\phi_i^{a r} = \sqrt{Z_i^a} \phi_i^a
\label{phi^r}
\ee
the part of the Lagrangian
(\ref{L_2}), describing the pions and kaons, takes the form
\be
L_\pi^{(2)} &=& \frac{1}{2} \left[ (P^2 - M^2_{\pi_1})~ \pi^2_1 +
2 \Gamma_\pi P^2~ \pi_1 \pi_2 + (P^2 - M^2_{\pi_2})~ \pi^2_2 \right],
\label{Lp}
\ee
\be
L_K^{(2)} &=& \frac{1}{2} [ (P^2 - M^2_{K_1} - \Delta^2)~ K^2_1
+ (P^2 - M^2_{K_2} - \Delta^2)~ K^2_2 \nonumber \\
&+& 2 \Gamma_K (P^2 - \Delta^2)~ K_1 K_2 ].
\label{LK}
\ee
Here
\be
\Gamma_a &=& \frac{\gamma_a}{\sqrt{Z_1^a Z_2^a}} =
\frac{I_2^{f a} \sqrt{Z}}{\sqrt{I_2^a I_2^{ff a} \bar{Z}}}.
\label{Gamma}
\ee
\be
M_{\pi_1}^2 &=& (4 Z I_2(m_u, m_u))^{-1}[\frac{1}{G_1}-8 I_1(m_u)] =
\frac{m_u^0}{4 Z m_u I_2 (m_u, m_u)}, \nonumber\\
M_{\pi_2}^2 &=& (4 I_2^{ff}(m_u, m_u))^{-1}[\frac{1}{G_1}-
8 I_1^{ff}(m_u)],
\label{Mp}
\ee
\be
M_{K_1}^2 &=& (4 Z I_2(m_u, m_s))^{-1}[\frac{1}{G_1} -
4 (I_1(m_u) + I_1(m_s))] + (Z^{-1} - 1) \Delta^2  \nonumber \\
&=& \frac{\frac{m_u^0}{m_u} + \frac{m_s^0}{m_s}}{4 Z
I_2(m_u, m_s)} + (Z^{-1} - 1) \Delta^2, \nonumber\\
M_{K_2}^2 &=& (4 I_2^{ff}(m_u, m_s))^{-1}[\frac{1}{G_1} -
4 (I_1^{ff}(m_u) + I_1^{ff}(m_s))].
\label{MK}
\ee
After the transformations of the meson fields ($sin \theta_a^0 =
\sqrt{\frac{1+\Gamma_a}{2}}$, see eq.(62))
\be
\phi^a
= cos( \theta_a - \theta_a^0) \phi_1^{ar}
- cos( \theta_a + \theta_a^0) \phi_2^{ar},   \nonumber \\
\phi^{'a}
= sin ( \theta_a - \theta_a^0) \phi_1^{ar}
- sin ( \theta_a + \theta_a^0) \phi_2^{ar}
\label{transf}
\ee
the Lagrangians (\ref{Lp}) and (\ref{LK}) take the diagonal forms
\be
L_\pi^{(2)} &=& \half (P^2 - M_\pi^2)~ \pi^2 +
\half (P^2 - M_{\pi'}^2)~ \pi^{' 2}, \\
L_K^{(2)} &=& \half (P^2 - M_K^2)~ K^2 +
\half (P^2 - M_{K'}^2)~ K^{' 2}.
\label{L_pK}
\ee
Here
\be
M^2_{(\pi, \pi')} = \frac{1}{2 (1 - \Gamma^2_\pi)}
[M^2_{\pi_1} + M^2_{\pi_2} \nonumber \\
(- , +)~ \sqrt{(M^2_{\pi_1} - M^2_{\pi_2})^2 +
(2 M_{\pi_1} M_{\pi_2} \Gamma_\pi)^2}], \\
M^2_{(K, K')} = \frac{1}{2 (1 - \Gamma^2_K)} [M^2_{K_1} + M^2_{K_2}
+ 2 \Delta^2 (1 - \Gamma^2_K)  \nonumber \\
(- , +)~ \sqrt{(M^2_{K_1} - M^2_{K_2})^2 +
(2 M_{K_1} M_{K_2} \Gamma_K)^2}].
\label{MpK}
\ee
and
\be
\tan 2 \bar{\theta_a} = \sqrt{\frac{1}{\Gamma^2_a} -
1}~\left[ \frac{M^2_{\phi_1^a}
- M^2_{\phi_2^a}}{M^2_{\phi_1^a} + M^2_{\phi_2^a}} \right],~~~~~~
2 \theta_a = 2 \bar{\theta_a} + \pi.
\label{tan}
\ee
In the chiral limit we obtain: $M_{\pi_1} = 0$, $M_{\pi_2}
\neq 0$ (see eqs. (\ref{Mp})) and
\be
M_\pi^2 &=& M_{\pi_1}^2 \; + \; {\cal O}(M_{\pi_1}^4 ),
\label{Mp_ch}
\ee
\be
M_{\pi'}^2 &=& \frac{M_{\pi_2}^2 + M_{\pi_1}^2 \Gamma_\pi}
{1 - \Gamma^2_\pi} \; + \; {\cal O}(M_{\pi_1}^4 ).
\label{Mp'_ch}
\ee
Thus, in the chiral limit the effective Lagrangian
eq.(\ref{Lp}) describes a massless Goldstone pion,
$\pi$, and a massive particle, $\pi'$. We obtained similar results
for the kaons.
\par
For the weak decay constants of the pions and kaons we obtain
(see \cite{volk_96})
\be
F_{\pi} &=& 2 m_u \sqrt{ Z I_2(m_u, m_u)}~
cos (\theta_\pi - \theta_\pi^0), \nonumber \\
F_{\pi'} &=& 2 m_u \sqrt{ Z I_2(m_u, m_u)}~
sin (\theta_\pi - \theta_\pi^0),
\label{f_p}
\ee
\be
F_{K} &=& (m_u + m_s) \sqrt{Z I_2(m_u, m_s)}~
cos (\theta_K - \theta_K^0), \nonumber \\
F_{K'} &=& (m_u + m_s) \sqrt{Z I_2(m_u, m_s)}~
sin (\theta_K - \theta_K^0).
\label{f_K}
\ee
In the chiral limit we have $\theta_a = \theta_a^0$
\be
sin \theta_a^0 = \sqrt{\frac{1 + \Gamma_a}{2}},\hspace{2cm}
cos \theta_a^0 = \sqrt{\frac{1 - \Gamma_a}{2}}
\label{theta_ch}
\ee
and
\be
F_\pi = \frac{m_u}{g_\pi},~~~F_K = \frac{(m_u + m_s)}{2 g_K},~~~
F_{\pi'} = 0,~~~F_{K'} = 0.
\label{f_ch}
\ee
Here we used eqs.(\ref{g_sigma}) and (\ref{g_phi}).
Therefore, in the chiral limit we obtain the Goldberger-Treimann
identities for the coupling constants $g_\pi$ and $g_K$.
The matrix elements of the divergences of the axial currents
between meson states and the vacuum equal (PCAC relations)
\be
\langle 0 | \partial^\mu A_\mu^a | \phi^b \rangle &=&
m_\phi^2 F_\phi \delta^{ab} ,
\label{A_phi} \\
\langle 0 | \partial^\mu A_\mu^a | \phi^{\prime\, b} \rangle &=&
m_{\phi'}^2 F_{\phi'} \delta^{ab}
\label{A_phi'} .
\ee
Then from eqs. (\ref{Mp_ch}) and (\ref{f_ch}) we can see that these
axial currents are conserved in the chiral limit, because their
divergences equal zero, according to the low-energy theorems.
\par
\section{The effective Lagrangian for the ground and excited
states of the vector mesons}
The free part of the effective Lagrangian (12) describing
the ground and excited states of the vector mesons has the form
\be
L^{(2)} (V) &=&
- \half \sum_{i, j = 1}^{2} \sum_{a = 0}^{8} V_i^{\mu a} (P)
R_{ij}^{\mu\nu a} (P) V_j^{\nu a}(P) ,
\label{LV_2}
\ee
where
\be
\sum_{a = 0}^{3} V_i^{\mu a} = (\omega_i^\mu)^2 +
(\rho_i^{0 \mu})^2 +
2 \rho_i^{+ \mu} \rho_i^{- \mu},~~~
(V_i^{4 \mu})^2 + (V_i^{5 \mu})^2 = 2 K_i^{* + \mu} K_i^{* - \mu}, \nonumber \\
(V_i^{6 \mu})^2 + (V_i^{7 \mu})^2 = 2 K_i^{* 0 \mu}
K_i^{* 0 \mu},~~~ (V_i^{8 \mu})^2 = (\phi_i^\mu)^2~~~~~~~~~~
\label{V^a}
\ee
and
\be
R_{ij}^{\mu \nu a} (P) =
-~ \frac{\delta_{ij}}{G_2} g^{\mu\nu}\hspace{5cm}
\nonumber \\
-~ i~ N_{\rm c} \; {\rm tr}\, \kint \left[
\frac{1}{\kslash + \half\Pslash - m_q^a}\gamma^\mu f_i^{a,V}
\frac{1}{\kslash - \half\Pslash - m_{q'}^a}  \gamma^\nu f_j^{a,V}
\right]  , \nonumber \\
f_1^{a,V} \equiv 1, \hspace{2em} f_2^{a,V} \;\; \equiv \;\; f_a^V ({\bf k}).\hspace{3cm}
\label{R_full}
\ee
To order $P^2$, one obtains
\be
R_{11}^{\mu\nu a} &=& W_1^a [P^2 g^{\mu\nu} - P^\mu P^\nu -
g^{\mu\nu} (\bar M^a_1)^2], \nonumber \\
R_{22}^{\mu\nu a} &=& W_2^a [P^2 g^{\mu\nu} - P^\mu P^\nu -
g^{\mu\nu} (\bar M^a_2)^2], \nonumber \\
R_{12}^{\mu\nu a} &=& R_{21}^{\mu\nu a} = \bar\gamma^a
[P^2 g^{\mu\nu} - P^\mu P^\nu - \frac{3}{2} \Delta^2 g^{\mu\nu}
\delta^{ab}|_{b = 4..7}].
\label{R_ij}
\ee
Here
\be
W_1^a &=& \frac{8}{3} I_2^a,~~~W_2^a = \frac{8}{3} I_2^{ff a},~~~
\bar\gamma^a = \frac{8}{3} I_2^{f a}, \\
(\bar M_1^a)^2 &=& (W_1^a G_2)^{-1} + \frac{3}{2}
\Delta^2 \delta^{ab}|_{b = 4..7}, \\
(\bar M_2^a)^2 &=& (W_2^a G_2)^{-1} + \frac{3}{2}
\Delta^2 \delta^{ab}|_{b = 4..7}.
\label{WM}
\ee
After renormalization of the meson fields
\be
V_i^{\mu a r} = \sqrt{W_i^a}~V_i^{\mu a}
\label{V^r}
\ee
we obtain the Lagrangians
\be
L_\rho^{(2)} &=& - \half [( g^{\mu\nu} P^2 - P^\mu P^\nu -
g^{\mu\nu} M^2_{\rho_1}) \rho^\mu_1 \rho^\nu_1 \nonumber \\
&+& 2 \Gamma_\rho  ( g^{\mu\nu} P^2 - P^\mu P^\nu) \rho_1^\mu
\rho_2^\nu + ( g^{\mu\nu} P^2 - P^\mu P^\nu -
g^{\mu\nu} M^2_{\rho_2}) \rho^\mu_2 \rho^\nu_2 ],
\label{L2_V1}
\ee
\be
L_\phi^{(2)} &=& - \half [( g^{\mu\nu} P^2 - P^\mu P^\nu -
g^{\mu\nu} M^2_{\phi_1}) \phi^\mu_1 \phi^\nu_1 \nonumber \\
&+& 2 \Gamma_\phi  ( g^{\mu\nu} P^2 - P^\mu P^\nu) \phi_1^\mu
\phi_2^\nu + ( g^{\mu\nu} P^2 - P^\mu P^\nu -
g^{\mu\nu} M^2_{\phi_2}) \phi^\mu_2 \phi^\nu_2 ],
\label{L2_V2}
\ee
\be
L_{K^*}^{(2)} &=& - \half [( g^{\mu\nu} P^2 - P^\mu P^\nu -
g^{\mu\nu} (\frac{3}{2} \Delta^2 + M^2_{K^*_1})) K^{*\mu}_1
K^{*\nu}_1 \nonumber \\
&+& 2 \Gamma_{K^*} ( g^{\mu\nu} P^2 - P^\mu P^\nu -
g^{\mu\nu} \frac{3}{2}
\Delta^2) K_1^{*\mu} K_2^{*\nu} \nonumber \\
&+& ( g^{\mu\nu} P^2 - P^\mu P^\nu -
g^{\mu\nu} (\frac{3}{2} \Delta^2 + M^2_{K^*_2})) K^{*\mu}_2 K^{*\nu}_2 ].
\label{L2_V3}
\ee
Here
\be
M_{\rho_1}^2 = \frac{3}{8 G_2 I_2(m_u, m_u)},~~~
M_{{K^*}_1}^2 = \frac{3}{8 G_2 I_2(m_u, m_s)},  \nonumber \\
M_{\phi_1}^2 = \frac{3}{8 G_2 I_2(m_s, m_s)},~~~
M_{\rho_2}^2 = \frac{3}{8 G_2 I^{ff}_2(m_u, m_u)}, \nonumber \\
M_{{K^*}_2}^2 = \frac{3}{8 G_2 I^{ff}_2(m_u, m_s)},~~~
M_{\phi_2}^2 = \frac{3}{8 G_2 I^{ff}_2(m_s, m_s)},
\label{MV_i}
\ee
\be
\Gamma_{a_{i,j}} = \frac{I_2^{f a}(m_i, m_j)}
{\sqrt{I_2^a(m_i, m_j)I_2^{ff a}(m_i, m_j)}}.
\label{GammaV}
\ee
After trasformations of the vector meson fields, similar to eqs.
(\ref{transf}) for the pseudoscalar mesons, the Lagrangians
(\ref{L2_V1},\ref{L2_V2},\ref{L2_V3}) take the diagonal form
\be
L^{(2)}_{V^a, \bar V^a} = - \half \left[ (g^{\mu\nu} P^2 -
P^\mu P^\nu - M^2_{V^a} ) V^{a \mu} V^{a \nu}  \right. \nonumber \\
\left. + (g^{\mu\nu} P^2 - P^\mu P^\nu -
M^2_{\bar V^a} ) \bar V^{a \mu} \bar V^{a \nu} \right],
\label{LDV}
\ee
where $V^a$ and $\bar V^a$ are the physical ground and excited
states vector mesons
\be
M^2_{\rho, \bar\rho} &=& \frac{1}{2 (1 - \Gamma^2_\rho)}~
\left[M^2_{\rho_1} + M^2_{\rho_2}~ (- , +)~ \sqrt{(M^2_{\rho_1} -
M^2_{\rho_2})^2 + (2 M_{\rho_1}M_{\rho_2} \Gamma_\rho)^2}
\right]  \nonumber \\
&=& M^2_{\omega, \bar\omega},
\label{Mrho}
\ee
\be
M^2_{\phi, \bar\phi} = \frac{1}{2 (1 - \Gamma^2_\phi)} \left[
M^2_{\phi_1} + M^2_{\phi_2}~ (- , +)~ \sqrt{(M^2_{\phi_1} -
M^2_{\phi_2})^2 + (2 M_{\phi_1}M_{\phi_2}
\Gamma_\phi)^2} \right] ,
\label{Mphi}
\ee
\be
M^2_{K^*, \bar K^*} = \frac{1}{2 (1 - \Gamma^2_{K^*})} \left[
M^2_{K^*_1} + M^2_{K^*_2} + 3 \Delta^2 (1 - \Gamma^2_{K^*})
\right.\nonumber \\
\left.
~( - , + )~ \sqrt{(M^2_{K^*_1} - M^2_{K^*_2})^2 +
(2 M_{K^*_1}M_{K^*_2} \Gamma_{K^*})^2} \right].
\label{MK^*}
\ee
\section{Numerical estimations}
We can now estimate numerically the masses of the pseudoscalar
and vector mesons and the weak decay constants $F_\pi$, $F_{\pi'}$,
$F_K$ and $F_{K'}$ in our model.
\par
Because the masses formulae and others equations ( for instance,
Goldber\-ger -- Treimann identity and so on) have new forms in the NJL
model with the excited states of mesons as compared with the usual
NJL model, where the excited states of mesons were ignored, the
values of basic parameters of this model ($m_u$, $m_s$, $\Lambda_3$,
$G_1$, $G_2$) could change. However, we see that one can use the
former values of the parameters $m_u$, $m_s$ and $\Lambda_3 = 1.03~ GeV$,
because the conditions (34) conserve the gap equation in the old form (16)
and one can satisfactory describe the decay $\rho \to 2 \pi$ in the new model
using the cut-off parameter $\Lambda_3 = 1.03~ GeV$ (compare with eq. (30)).
In new model $G_1 = 3.469 GeV^{-2}$. It is very close to former value
$G_1 = 3.48~ GeV^{-2}$, because $M_\pi \approx M_{\pi_1}$ (see eqs. (50),
(55) and (23)). For the coupling
constant $G_2$ the new value $G_2 = 12.5~ GeV ^{-2}$ will
be used, which more noticeably differs from the former
value $G_2 = 16~ GeV^{-2}$ (see section 2). It is a
consequence of the fact that the mass $M_{\rho_1}$ noticeably
differs from the physical mass $M_\rho$ of the ground state $\rho$
(see eqs. (\ref{MV_i}) and (\ref{Mrho})).
\par
Using these basic parameters and the internal form factor
parameter $d_{uu} = -1.784~ GeV^{-2}$ (see eq. (\ref{d_a}))
and choosing the external form factor parameters
$c_{uu}^\pi = 1.37$ and $c_{uu}^\rho = 1.26$, one finds
\be
M_\rho &=& 768.3~ MeV,~~~ M_{\rho'} = 1.49~ GeV,  \nonumber \\
M_\pi &=& 136~ MeV,~~~M_{\pi'} = 1.3~ GeV.
\label{Mprot}
\ee
\be
\Gamma_\pi = 0.474,~~~\Gamma_\rho = 0.545.
\label{GG}
\ee
$\Gamma_\pi = \sqrt{\frac{Z}{\bar{Z}}}~\Gamma_\rho$
( see eqs. (\ref{Gamma}) and
(\ref{GammaV})). The experimental values are equal to \cite{Rev_96}
\be
M^{exp}_\rho &=& 768.5 \pm 0.6~ MeV,~~~ M^{exp}_{\rho'} =
1465 \pm 25~ MeV, \nonumber \\
M^{exp}_{\pi^+} &=& 139.57~ MeV,~~~M^{exp}_{\pi^0} = 134.98~ MeV, \nonumber \\
M^{exp}_{\pi'} &=& 1300 \pm 100~ MeV.
\label{Mproe}
\ee
>From eq. (\ref{f_p}), one obtains
\be
F_\pi = 93~ MeV,~~~~  F_{\pi'} = 0.57~ MeV, \nonumber \\
\frac{F_{\pi'}}{F_\pi} \approx \sqrt{\frac{1}{\Gamma_\pi^2 - 1}}~
(\frac{M_\pi}{M_{\pi'}})^2
\label{ff'}
\ee
Using the internal form factor parameter $d_{us} = -1.757~ GeV^{-2}$
(see eq. (\ref{d_a})) and choosing the external form factor parameters
$c_{us}^K = 1.45$, $c_{us}^{K^*} = 1.5$, one finds
\be
M_{K^*} &=& 887~ MeV,~~~M_{K^{*'}}= 1479~ MeV, \nonumber \\
M_K &=& 496~ MeV,~~~M_{K'} = 1450~ MeV,
\label{MKK't}
\ee
\be
\Gamma_K = 0.412,~~~\Gamma_{K^*} = 0.473.
\label{GGK}
\ee
The experimental values are equal to
\be
M^{exp}_{K^*} &=& 891.59 \pm 0.24~ MeV,~~~ M^{exp}_{K^{*'}} =
1412 \pm 12~ MeV, \nonumber \\
M^{exp}_{K^+} &=& 493.677 \pm 0.016~ MeV,~~~
M^{exp}_{K^0} = 497.672 \pm 0.031~ MeV,   \nonumber \\
M^{exp}_{K'} &=& 1460~ MeV (?).
\label{MKK'e}
\ee
>From the eq. (\ref{f_K}), one gets
\be
F_K = 1.16 F_\pi = 108~ MeV, \hspace{3cm} F_{K'} = 3.3~ MeV.
\label{fKK'}
\ee
And for the $\phi$ and $\phi'$ we obtain, using the form
factor parameters $d_{ss} = -1.727~ GeV^{-2}$ (see eq. (\ref{d_a}))
and $c_{ss}^{\phi} = 1.41$
\be
M_{\phi} = 1019~ MeV,~~~M_{\phi'} = 1682~ MeV,~~~\Gamma_{\phi} = 0.411
\label{phit}
\ee
The experimental values are equal to
\be
M^{exp}_{\phi} = 1019.413 \pm 0.008~ MeV,~~~M^{exp}_{\phi'} =
1680 \pm 50~ MeV.
\label{phie}
\ee
We can see that the parameters $c_{us}$ and $c_{ss}$ are close
to each other.
\section{Summary and conclusions}
Let us discuss the obtained results.
Conditions (34) allow us to conserve the gap equations in the
form usual for the standard NJL model (see eqs.(16)) and to fix all the
internal form factor parameters $d_a$. These parameters are equal to
each other with accuracy $1.5{\%}$. As a result, the constituent quark
masses $m_u$ and $m_s$ and the cut-off parameter $\Lambda_3$ conserve
their former values after introducing excited meson states into the model.
(The constant $\bar{g}_\rho$ describing the decay $\rho \to 2~\pi$
in the new model approximately equals the former constant $g_\rho$ (eq.(30))
for $\Lambda_3 = 1.03 GeV$.)
\par
We can describe all the excited states of strange mesons (pseudoscalar
kaons, vector kaons and the $\phi$-meson) using practically only one value of
the external parameter $c_a$, because $c_{us}^{K^*} \approx c_{us}^K
\approx c_{ss}^{\phi}$ with accuracy $3{\%}$.
\par
A more complicated situation takes place in the sector of light mesons
consisting of $u$ and $d$ quarks. Here the parameters $c_{uu}^\pi$ and
$c_{uu}^\rho$ not only differ noticeably from similar parameters of
strange mesons but they are also not equal to each other. This difference
equals $8{\%}$. The coupling constant of the effective
four-quark interaction describing the excited $\rho$-meson states,
$G_2 (c_{uu}^\rho)^2$, is 1.4 times less than the coupling constants
$G_2 c_{us}^2$ and $G_2 c_{ss}^2$ describing the excited states of strange
mesons.
\par
The mass of
the first radial excited state of the pion has an interesting
history. Three years ago the new experimental information
about excited states in the few-GeV region, e.g on the
$\pi'$ meson, was obtained at IHEP (Protvino). Indications
of the light resonance in diffractive
production of $3 \pi$--states have lead to speculations that the
mass of the $\pi'$  may be considerably lower at $\approx 750 MeV$
\cite{ivanshin_93}. To describe this pion state in our model,
it is necessary to use a much larger value of $c_{uu}^\pi = 1.67$.
And for this pion state the decay channel $\pi' \to \rho \pi$
observed in experiment \cite{Rev_96} is forbidden. Therefore,
we prefer to consider the first radial excited pion state as
a state with mass 1.3 GeV.
\par
Besides the description of the excited meson masses, the weak decay
coupling constants $F_{\pi'}$, $F_{K}$ and $F_{K'}$ were calculated.
In future we are going to calculate the decay widths of the excited meson
nonets. The preliminary result on the decay width $\pi' \to \rho \pi$
satisfies the experimantal data.
\par
We have here considered the simplest extension of the
NJL--model with polynomial meson--quark form factors and
have shown that this model can be useful for describing the excited
states of mesons.
\par
The author would like to thank Drs. S.B.Gerasimov and C.Weiss for
the fruitful discussions and Prof.A.Di-Giacomo for the kind
hospitality in Istituto Nazionale di Fisica Nucleare (Pisa),
where the main part of this work was performed.
%

\end{document}